%% file: main.tex
\pdfoutput=1

\documentclass[12pt,a4paper]{article}

\usepackage[utf8]{inputenc}
\usepackage[T1]{fontenc}
\usepackage{lmodern}

\usepackage{ifthen} 
\newboolean{pdflatex}
\setboolean{pdflatex}{true} 

\newboolean{articletitles}
\setboolean{articletitles}{true} 

\newboolean{uprightparticles}
\setboolean{uprightparticles}{false} 

\input{preamble}

\usepackage{ctable}
\setupctable{mincapwidth=\textwidth}

\newcommand{\Rxy}{\ensuremath{R_{xy}}}
\newcommand{\vpion}{\ensuremath{\pi_{\text{v}}}}

\newcommand{\Ztomumu}{\ensuremath{\Z\rightarrow\mumu}}
\newcommand{\Htopivpiv}{{\ensuremath{\Higgs{}\rightarrow\vpion\vpion}}}

\RequirePackage{subcaption}
\usepackage{csquotes}

\usepackage[style=lhcbbibstyle,mcite]{biblatex}

\begin{document}

\renewcommand{\thefootnote}{\fnsymbol{footnote}}
\setcounter{footnote}{1}

\input{title-LHCb-PAPER}


\renewcommand{\thefootnote}{(\alph{footnote})}
\setcounter{footnote}{0}

\clearpage%



\pagestyle{plain} 
\setcounter{page}{1}
\pagenumbering{arabic}

\newlength{\fullfigwidth}\setlength{\fullfigwidth}{15cm}
\newlength{\halffigwidth}\setlength{\halffigwidth}{7.5cm}


\input{introduction}

\input{detector}

\input{selection}

\input{systematics}

\input{results}

\input{acknowledgements}

\clearpage
\printbibliography%

\newpage
\input{LHCb_Authorship_flat_20-Dec-2016.tex}

\end{document}

%% file: preamble.tex
\usepackage[top=1in, bottom=1.25in, left=1in, right=1in]{geometry}

\usepackage{microtype}
\usepackage{lineno}  
\usepackage{caption} 

\usepackage{graphicx}  
\usepackage{color}
\usepackage{colortbl}
\graphicspath{{./figs/}} 

\usepackage{amsmath} 
\usepackage{amssymb}
\usepackage{amsfonts}
\usepackage{upgreek} 

\newcommand*\patchAmsMathEnvironmentForLineno[1]{%
\expandafter\let\csname old#1\expandafter\endcsname\csname #1\endcsname
\expandafter\let\csname oldend#1\expandafter\endcsname\csname
end#1\endcsname
 \renewenvironment{#1}%
   {\linenomath\csname old#1\endcsname}%
   {\csname oldend#1\endcsname\endlinenomath}%
}
\newcommand*\patchBothAmsMathEnvironmentsForLineno[1]{%
  \patchAmsMathEnvironmentForLineno{#1}%
  \patchAmsMathEnvironmentForLineno{#1*}%
}
\AtBeginDocument{%
\patchBothAmsMathEnvironmentsForLineno{equation}%
\patchBothAmsMathEnvironmentsForLineno{align}%
\patchBothAmsMathEnvironmentsForLineno{flalign}%
\patchBothAmsMathEnvironmentsForLineno{alignat}%
\patchBothAmsMathEnvironmentsForLineno{gather}%
\patchBothAmsMathEnvironmentsForLineno{multline}%
\patchBothAmsMathEnvironmentsForLineno{eqnarray}%
}

\usepackage{hyperref}    
\usepackage[all]{hypcap} 

\usepackage[per-mode=symbol,range-phrase=\text{--},range-units=single,list-units=single,separate-uncertainty=true,detect-shape]{siunitx}
\input{lhcb-symbols-def} 

\usepackage[capitalise]{cleveref}
\crefname{section}{Sect.}{Sections}

\usepackage{adjustbox}

%% file: lhcb-symbols-def.tex
\def\lhcb{\mbox{LHCb}}
\def\atlas{\mbox{ATLAS}}
\def\cms{\mbox{CMS}}

\def\velo{VELO}

\def\spd{SPD}

\def\MagUp{\mbox{\emph{Mag\kern -0.05em Up}}} 

\def\lone{L0}

\def\hltone{HLT1}
\def\hlttwo{HLT2}

\ifthenelse{\boolean{uprightparticles}}%
{

 \def\Pmu{\ensuremath{\upmu}}

 \def\Ppsi{\ensuremath{\uppsi}}

 \def\PDelta{\ensuremath{\Delta}}                 
 \def\PXi{\ensuremath{\Xi}}                 
 \def\PLambda{\ensuremath{\Lambda}}                 
 \def\PSigma{\ensuremath{\Sigma}}                 
 \def\POmega{\ensuremath{\Omega}}                 
 \def\PUpsilon{\ensuremath{\Upsilon}}                 
 
 \def\PB{\ensuremath{\mathrm{B}}}                 
                  
 \def\PD{\ensuremath{\mathrm{D}}}

 \def\PH{\ensuremath{\mathrm{H}}}                 
                  
 \def\PJ{\ensuremath{\mathrm{J}}}                 
 \def\PK{\ensuremath{\mathrm{K}}}

 \def\PZ{\ensuremath{\mathrm{Z}}}                 
                  
 \def\Pb{\ensuremath{\mathrm{b}}}                 
 \def\Pc{\ensuremath{\mathrm{c}}}

 \def\P_i{\ensuremath{\mathrm{i}}}

 \def\Pq{\ensuremath{\mathrm{q}}}                 
                  
 \def\Ps{\ensuremath{\mathrm{s}}}

}
{

 \def\Pmu{\ensuremath{\mu}}

 \def\Ppsi{\ensuremath{\psi}}                 
                  
 \mathchardef\PDelta="7101   
 \mathchardef\PXi="7104      
 \mathchardef\PLambda="7103  
 \mathchardef\PSigma="7106   
 \mathchardef\POmega="710A   
 \mathchardef\PUpsilon="7107 
                  
 \def\PB{\ensuremath{B}}                 
                  
 \def\PD{\ensuremath{D}}

 \def\PH{\ensuremath{H}}                 
                  
 \def\PJ{\ensuremath{J}}                 
 \def\PK{\ensuremath{K}}

 \def\PZ{\ensuremath{Z}}                 
                  
 \def\Pb{\ensuremath{b}}                 
 \def\Pc{\ensuremath{c}}

 \def\P_i{\ensuremath{i}}

 \def\Pq{\ensuremath{q}}                 
                  
 \def\Ps{\ensuremath{s}}

}

\makeatletter
\ifcase\@ptsize\relax
  \newcommand{\miniscule}{\@setfontsize\miniscule{4}{5}}
\or
  \newcommand{\miniscule}{\@setfontsize\miniscule{5}{6}}
\or
  \newcommand{\miniscule}{\@setfontsize\miniscule{5}{6}}
\fi
\makeatother

\DeclareRobustCommand{\optbar}[1]{\shortstack{{\miniscule(\rule[.5ex]{1.25em}{.18mm})}
  \\ [-.7ex] $#1$}}

\def\mumu{{\ensuremath{\Pmu^+\Pmu^-}}}

\def\Higgs{{\ensuremath{\PH^0}}}

\def\Z{{\ensuremath{\PZ}}}

\def\quark{{\ensuremath{\Pq}}}
\def\quarkbar{{\ensuremath{\overline \quark}}}
\def\qqbar{{\ensuremath{\quark\quarkbar}}}

\def\squark{{\ensuremath{\Ps}}}
\def\squarkbar{{\ensuremath{\overline \squark}}}
\def\ssbar{{\ensuremath{\squark\squarkbar}}}
\def\cquark{{\ensuremath{\Pc}}}
\def\cquarkbar{{\ensuremath{\overline \cquark}}}
\def\ccbar{{\ensuremath{\cquark\cquarkbar}}}
\def\bquark{{\ensuremath{\Pb}}}
\def\bquarkbar{{\ensuremath{\overline \bquark}}}
\def\bbbar{{\ensuremath{\bquark\bquarkbar}}}

\def\kaon{{\ensuremath{\PK}}}

\def\KorKbar{\kern 0.18em\optbar{\kern -0.18em K}{}} 

\def\Kstarz{{\ensuremath{\kaon^{*0}}}}

  \def\Dbar{{\kern 0.2em\overline{\kern -0.2em \PD}{}}} 

\def\DorDbar{\kern 0.18em\optbar{\kern -0.18em D}{}} 

\def\B{{\ensuremath{\PB}}}

\def\BorBbar{\kern 0.18em\optbar{\kern -0.18em B}{}} 
\def\Bz{{\ensuremath{\B^0}}}

\def\Bd{{\ensuremath{\B^0}}}

\def\jpsi{{\ensuremath{{\PJ\mskip -2mu/\mskip -2mu\Ppsi\mskip 2mu}}}}

  \def\Y#1S{\ensuremath{\PUpsilon{(#1S)}}}

\def\Lbar{{\ensuremath{\kern 0.1em\overline{\kern -0.1em\PLambda}}}} 
\def\LorLbar{\kern 0.18em\optbar{\kern -0.18em \PLambda}{}} 

\def\BF{\mathcal{B}}

\def\BR{\BF}
\newcommand{\decay}[2]{\ensuremath{#1\!\to #2}}         

\def\to{\ensuremath{\rightarrow}}

\def\BdToJPsiKst{\decay{\Bd}{\jpsi\Kstarz}}

\def\FL{\ensuremath{F_{\mathrm{L}}}}

\RequirePackage{siunitx}
\DeclareSIUnit\degree{°}
\DeclareSIUnit\clight{\ensuremath{\textit{c}}}
\DeclareSIUnit{\electronvolt}{e\kern -0.1em V\!} 
\DeclareSIUnit\GeVc{\giga\electronvolt\per\clight}
\DeclareSIUnit\GeVcc{\giga\electronvolt\per\kern -0.1em\square\clight}
\DeclareSIUnit\MeVc{\mega\electronvolt\per\clight}
\DeclareSIUnit\MeVcc{\mega\electronvolt\per\square\clight}
\DeclareSIUnit\GeVGeVcccc{\giga\electronvolt\squared\per\clight\tothe{4}}

\DeclareSIUnit\nb{\nano\barn}
\DeclareSIUnit\pb{\pico\barn}
\DeclareSIUnit\fb{\femto\barn}
\DeclareSIUnit\invnb{\nb\tothe{-1}}
\DeclareSIUnit\invpb{\pb\tothe{-1}}
\DeclareSIUnit\invfb{\fb\tothe{-1}}

\def\gsim{{~\raise.15em\hbox{$>$}\kern -.85em 
          \lower.35em\hbox{$\sim$}~}}
\def\lsim{{~\raise.15em\hbox{$<$}\kern -.85em 
          \lower.35em\hbox{$\sim$}~}}

\def\ptot{\mbox{$p$}}
\def\pt{\mbox{$p_\mathrm{T}$}}

\def\geant{\mbox{\textsc{Geant4}}}

\def\pythia{\mbox{\textsc{Pythia}}}

\def\run1{Run~1}

\def\tell1{TELL1}
\def\ukl1{UKL1}



%% file: title-LHCb-PAPER.tex

\begin{titlepage}
\pagenumbering{roman}

\vspace*{-1.5cm}
\centerline{\large EUROPEAN ORGANIZATION FOR NUCLEAR RESEARCH (CERN)}
\vspace*{1.5cm}
\noindent
\begin{tabular*}{\linewidth}{lc@{\extracolsep{\fill}}r@{\extracolsep{0pt}}}
\ifthenelse{\boolean{pdflatex}}
{\vspace*{-2.7cm}\mbox{\!\!\!\includegraphics[width=.14\textwidth]{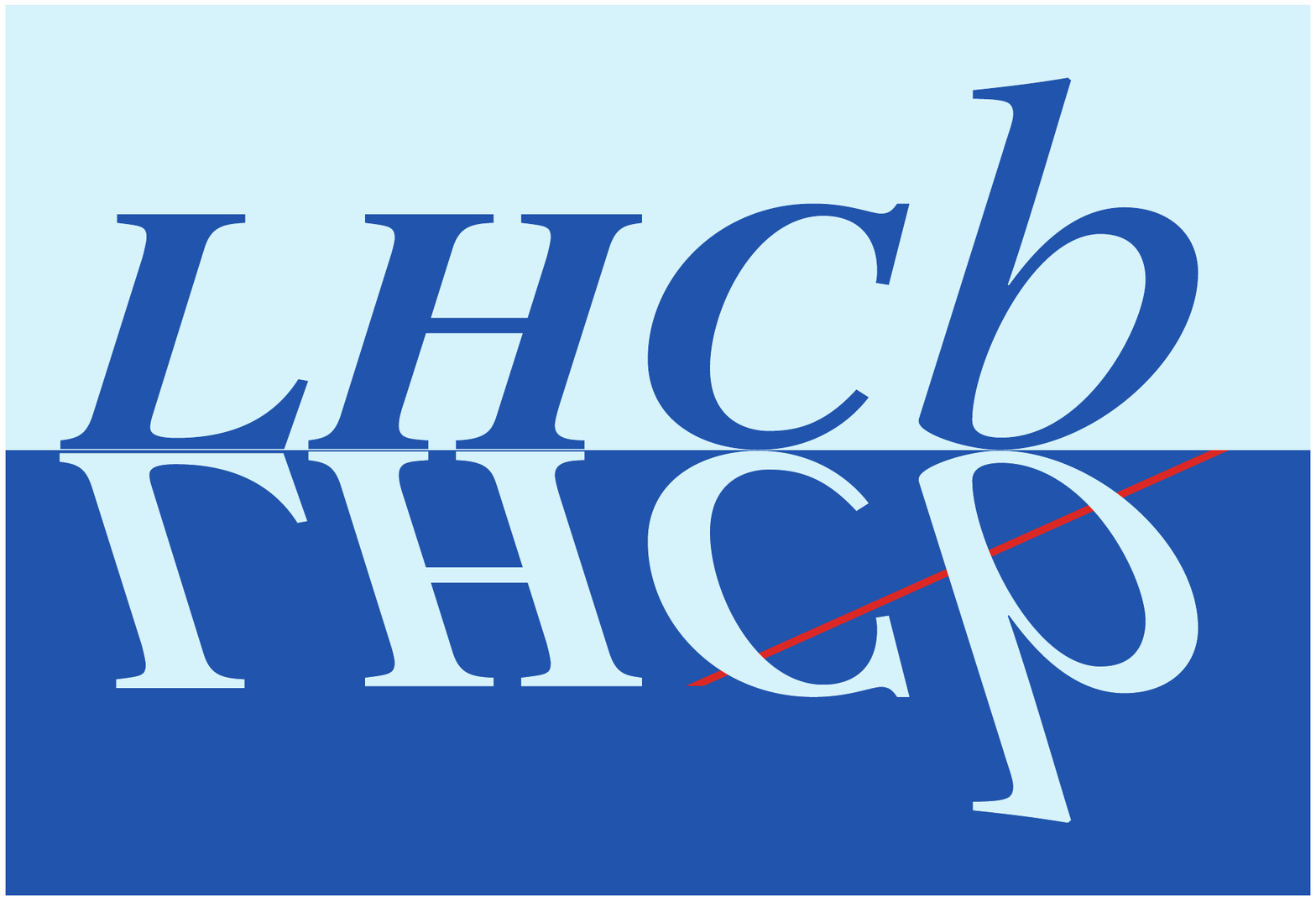}} & &}%
{\vspace*{-1.2cm}\mbox{\!\!\!\includegraphics[width=.12\textwidth]{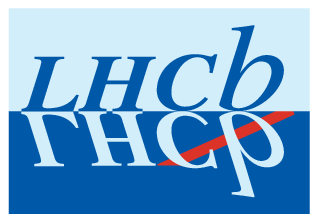}} & &}%
\\
 & & CERN-EP-2017-083 \\  
 & & LHCb-PAPER-2016-065 \\  
 & & December 3rd, 2017 \\ 
 & & \\
\end{tabular*}

\vspace*{2.0cm}

{\normalfont\bfseries\boldmath\huge
\begin{center}
  Updated search for long-lived particles decaying to jet pairs
\end{center}
}

\vspace*{1.0cm}

\begin{center}
The LHCb collaboration\footnote{Authors are listed at the end of this paper.}
\end{center}

\vspace{\fill}

\begin{abstract}
  \noindent
  A search is presented for long-lived particles with a mass between
  \SIlist{25;50}{\GeVcc} and a lifetime between \SIlist{2;500}{\ps},
  using proton-proton collision data corresponding to an integrated
  luminosity of \SI{2.0}{\invfb}, collected by the \lhcb{} detector at
  centre-of-mass energies of \SIlist{7;8}{TeV}.
  The particles are assumed to be pair-produced in
  the decay of a \SI{125}{\GeVcc} Standard-Model-like Higgs boson. The
  experimental signature is a single long-lived particle, identified
  by a displaced vertex with two associated jets. No excess above
  background is observed and limits are set on the production
  cross-section as a function of the mass and
  lifetime of the long-lived particle.
\end{abstract}

\vspace*{1.0cm}

\begin{center}
  Published in \href{https://doi.org/10.1140/epjc/s10052-017-5178-x}{Eur.~Phys.~J.~C77 (2017) 812}
\end{center}

\vspace{\fill}

{\footnotesize 
\centerline{\copyright~CERN on behalf of the \lhcb{} collaboration, licence \href{http://creativecommons.org/licenses/by/4.0/}{CC-BY-4.0}.}}
\vspace*{2mm}

\end{titlepage}


\newpage
\setcounter{page}{2}
\mbox{~}
%
%
%
%

\cleardoublepage

%% file: introduction.tex
\section{Introduction}
\label{sec:introduction}

Various extensions of the Standard Model (SM) feature new particles
whose couplings to lighter states are sufficiently small to result in
detectable lifetimes. In this paper we report on a search for
such long-lived particles, which are assumed to be pair-produced in the decay of
a Standard-Model-like Higgs boson, and subsequently decay into
a quark-antiquark pair.
Such a signature is present in models with a hidden-sector non-Abelian
gauge group, where the Standard Model Higgs boson acts as
a portal~\cite{StrasslerZurekHV,StrasslerZurekHVHiggs,Chang:2008cw,Craig:2015pha,Curtin:2015fna}. The
new scalar particle represents the lightest state in the hidden sector
and is called a hidden-valley pion (\vpion{}) throughout this paper.
Experimental constraints on the properties of the Higgs boson of mass
\SI{125}{\GeVcc} observed by the \atlas{} and \cms{}
collaborations~\cite{ATLASHiggsDiscovery,CMSHiggsDiscovery} still
allow for branching fractions of non-SM decay modes of up to
30\%~\cite{ATLASCMSRun1HiggsCombination}.

Data collected with the LHCb experiment in 2011 and 2012 are used for
this analysis, restricted to periods in which suitable triggers were
available. The data sample analysed corresponds to \SI{0.62}{\invfb}
at a centre-of-mass energy of $\sqrt{s}=\SI{7}{TeV}$ and \SI{1.38}{\invfb} at
$\sqrt{s}=\SI{8}{TeV}$.  In simulated events with \vpion{} pairs originating from
a Higgs boson decay it is found that in most cases no more than one of the
two \vpion{} decays occurs inside the LHCb acceptance. Consequently,
the experimental signature is a single \vpion{} particle. The
candidate is identified by its decay to two hadronic jets originating
from a displaced vertex, with a transverse distance to the
proton-proton collision axis (\Rxy{}) of at least \SI{0.4}{mm}. The
vertex is required to have at least five tracks reconstructed in the
LHCb vertex detector. The analysis is sensitive to \vpion{} particles
with a mass between 25 and \SI{50}{\GeVcc} and a lifetime between 2
and \SI{500}{ps}.  The lifetime range is limited due to the presence
of large prompt backgrounds at short decay times and the acceptance of
the vertex detector for long decay times. The lower boundary on the
mass range arises from the requirement to identify two hadronic jets
while the upper boundary is driven by the geometric acceptance of the
detector.

This paper presents an update of an earlier analysis, which considered only the data set corresponding to an integrated luminosity of \SI{0.62}{\invfb} collected at $\sqrt{s}=7$~TeV~\cite{LHCb-PAPER-2014-062}. Similar searches for hidden-valley
particles decaying to jet pairs were performed by the
D0~\cite{D0DVHiddenValleyHiggs2009},
CDF~\cite{CDFDVHiddenValleyHiggs2011},
ATLAS~\cite{ATLASDVEXO2011,ATLASDVEXO2012,ATLASDVEXO2012CalRatio} and
CMS~\cite{CMSDVDiJet2012} collaborations. Compared to these analyses, this search is
sensitive to \vpion{} particles with relatively low mass and
lifetime. The LHCb collaboration has also performed a search for
events with two displaced high-multiplicity
vertices~\cite{LHCb-PAPER-2016-014} and a search for events with a
lepton from a high-multiplicity displaced
vertex~\cite{LHCb-PAPER-2016-047} in the context of SUSY models, and
several searches for so far unknown long-lived particles in $B$-meson
decays~\cite{LHCb-PAPER-2013-064,LHCb-PAPER-2011-038,LHCb-PAPER-2015-036,LHCb-PAPER-2016-052}.

%% file: detector.tex
\section{Detector and event simulation}
\label{sec:Detector}

The \lhcb{} detector~\cite{Alves:2008zz,LHCb-DP-2014-002} is a
single-arm forward spectrometer covering the \mbox{pseudorapidity}
range $2<\eta <5$, designed for the study of particles containing
\bquark{} or \cquark{} quarks. The detector includes a high-precision
tracking system consisting of a silicon-strip vertex detector (\velo{})
surrounding the $pp$ interaction region, a large-area silicon-strip
detector located upstream of a dipole magnet with a bending power of
about $4{\mathrm{\,Tm}}$, and three stations of silicon-strip
detectors and straw drift tubes placed downstream of the magnet.  The
tracking system provides a measurement of the momentum, \ptot, of charged
particles with a relative uncertainty that varies from 0.5\% at low
momentum to 1.0\% at \SI{200}{\GeVc}.  The minimum distance of a track to a
primary vertex (PV), the impact parameter (IP), is measured with a
resolution of $(15+(\SI{29}{\GeVc})/\pt)\;\si{\micro\meter}$, where \pt{} is the component of the
momentum transverse to the collision axis.  Different types of
charged hadrons are distinguished using information from two
ring-imaging Cherenkov detectors.  Photons, electrons and hadrons are
identified by a calorimeter system consisting of scintillating-pad (\spd{}) and
preshower detectors, an electromagnetic calorimeter and a hadronic
calorimeter. Muons are identified by a system composed of alternating
layers of iron and multiwire proportional chambers.

The model for the production of \vpion{} particles through the Higgs
portal is fully specified by three parameters: the mass of the Higgs
boson and the mass and lifetime of the \vpion{}. The Higgs boson mass
is taken to be \SI{125}{\GeVcc}, and its production through the
gluon-gluon fusion process is simulated with the \pythia{}8
generator~\cite{Sjostrand:2007gs}, with a specific \lhcb{}
configuration~\cite{LHCb-PROC-2010-056} and using the CTEQ6
leading-order set of parton density functions~\cite{cteq6}.
The interaction of the generated particles with the detector, and its
response, are implemented using the \geant{}
toolkit~\cite{Allison:2006ve,Agostinelli:2002hh} as described
in Ref.~\cite{LHCb-PROC-2011-006}. Signal samples with \vpion{} masses of
\SIlist{25;35;43;50}{\GeVcc} and lifetimes of
\SIlist{10;100}{\pico\second} are generated. In the simulated events
the long-lived particles decay exclusively as $\vpion\to\bbbar$, since
this decay mode is generally preferred in the Higgs portal
model. Samples with decays to \cquark{}- and \squark{}-quark pairs are
generated as well, but only in the scenario with a mass of
\SI{35}{\GeVcc} and a lifetime of \SI{10}{\pico\second}.

%% file: selection.tex
\section{Event selection}
\label{sec:selection}

The experimental signature for this analysis is a single displaced
vertex with two associated jets.  Only decays that produce a
sufficient number of tracks in the \velo{} for a vertex to be
reconstructed are considered.  Due to the geometry of the vertex
detector, this restricts the sample to decay points up to about
\SI{200}{\milli\meter} from the nominal interaction point along the
beam direction, and up to about \SI{30}{\milli\meter} in the transverse
direction, thereby limiting the decay time acceptance. The selection
strategy is the same as used in the analysis of
Ref.~\cite{LHCb-PAPER-2014-062}. Reconstructed tracks are used to find
the decay vertex, and jets are built out of reconstructed
particles compatible with originating from that vertex.  Constraints
on the signal yield are determined from a fit to the dijet invariant
mass distribution. The main source of background is displaced
vertices from heavy-flavour decays or interactions of particles with
detector material. To take into account the strong dependence of the
background level on the separation from the beam axis, different
selection criteria are used in different bins of \Rxy{}, and the final
fit is performed in bins of this variable.

The selection consists of online (trigger) and offline parts. The
trigger~\cite{LHCb-DP-2012-004} is divided into a hardware (L0) and a
software (HLT) stage.  The L0 requires a muon with high \pt{} or a
hadron, photon or electron with high transverse energy in the
calorimeters. In order to reduce the processing time of the subsequent
trigger stages, events with a large hit multiplicity in the \spd{} are
discarded. The software stage is divided into two parts, which for
this analysis differ between the 2011 and 2012 data. In the 2011
sample, the first software stage (HLT1) requires a single high-\pt{}
track with a large impact parameter. The HLT1 selection for the 2012
sample was complemented with a two-track vertex signature with looser
track quality criteria, in order to improve the efficiency at large
displacements. At the second stage of the software trigger (HLT2),
events are required to pass either a dedicated inclusive displaced-vertex
selection or a standard topological $B$ decay selection, which
requires a two-, three- or four-track vertex with a significant
displacement from all PVs~\cite{LHCb-DP-2012-004}. The inclusive displaced-vertex selection uses an
algorithm similar to that used for the LHCb primary vertex
reconstruction~\cite{LHCb-PUB-2014-044}. A combination of requirements
on the minimum number of tracks in the vertex (at least four), the
distance \Rxy{} of the vertex to the beam axis (at least
\SI{0.4}{\milli\meter}), the invariant mass of the particles associated with the vertex (at least
\SI{2}{\GeVcc}) and the scalar sum \pt{} of the tracks that form the
vertex (at least \SI{3}{\GeVc}), is used to define a set of
trigger selections with sufficiently low rate.

Before the offline selection can be applied, the displaced vertex
corresponding to the decay of the \vpion{} candidate must be
reconstructed.  For those events in which the \hlttwo{} inclusive displaced-vertex
selection was successful, the same vertex candidate found in the
trigger is used; this approach differs from that used in the previous
\lhcb{} analysis~\autocite{LHCb-PAPER-2014-062} and simplifies the
evaluation of systematic uncertainties.  For events selected only by the topological $B$ trigger, 
a modified
version of the algorithm is run on the output of the offline
reconstruction with the following criteria: vertices with
\(0.4 < \Rxy{} < \SI{1}{\milli\meter}\) must have at least eight
tracks and the invariant mass of the system must exceed
\SI{10}{\GeVcc}, vertices with \(1 < \Rxy{} < \SI{5}{\milli\meter}\)
must have at least six tracks, and those with
\(\Rxy{} > \SI{5}{\milli\meter}\) must have at least five tracks.  To
exclude background due to interactions with the detector material,
vertices inside a veto region around the \velo{} detector elements are
discarded.  Events with many parallel displaced tracks, which can
arise from machine background, are identified by the azimuthal
distribution of hits in the \velo{} and are also discarded.

Next, jets are reconstructed following a particle flow
  approach.  The same set of inputs as in
Ref.~\autocite{LHCb-PAPER-2013-058} is used, namely tracks of charged
particles and calorimeter energy deposits, after subtraction of the
energy associated with charged particles. 
To remove background, tracks that are compatible with coming from a
PV, tracks with a smaller impact parameter to any primary vertex than
to the displaced vertex, and tracks that have an impact parameter to
the displaced vertex larger than \SI{2}{\milli\meter} are all discarded.
The anti-\(k_T\) jet clustering algorithm is used~\autocite{antikt},
with a distance parameter of \(R=0.7\).
The jet momentum and jet mass are calculated from the
  four-vectors of all constituents of the jet. 
In simulated events the jet energy response is found to be close to
unity except for the lowest jet momenta, near the minimally required
  transverse momentum of \(\SI{5}{\GeVc}\).  Therefore, no jet energy
correction was applied for this search.

To enhance the jet purity the fraction of the jet energy carried by
charged particles should be at least 0.1, there should be at least one
track with transverse momentum above \SI{0.9}{\GeVc}, no pair of
constituents should carry 90\% of the jet energy, and no single
charged or neutral constituent should contribute more than 70\% or
50\% of the total energy, respectively.
To ensure that they can reliably be associated to a vertex, the jets
are also required to have at least two constituents with track
segments in the VELO. 
To account for differences in trigger and background
  conditions, for the 2012 data this requirement was
tightened to at least four segments for $\Rxy<\SI{1}{\milli\meter}$,
and at least three segments for
\(1 < \Rxy < \SI{2}{\milli\meter}\).
For each jet an origin point is reconstructed from the jet
constituents with \velo{} information.  The jet trajectory is defined
based on this origin point and the momentum of the jet.  Any jet whose
trajectory does not point back to the candidate vertex within
\SI{2}{\milli\meter}, or points more closely to a primary vertex, is
removed.
Only candidates with at least two jets passing these criteria are retained.

Two final criteria are applied to the dijet candidates.
The first is that the momentum vector of the dijet candidate should be aligned with the
displacement vector from a PV to the reconstructed vertex position.
This is implemented as a requirement on the dijet invariant mass
divided by the corrected mass, \(m/m_\text{corr} > 0.7\). The
corrected mass is computed as
\(m_\text{corr} = \sqrt{m^2+\left(p\sin\theta\right)^2} +
p\sin\theta\)~\cite{corrmass}, where \(m\) and \(p\) are
the reconstructed mass and momentum of the dijet, and $\theta$ is the minimum
angle between the momentum vector and the displacement vectors to the vertex from any
PV in the event.  A requirement on \(m/m_\text{corr}\) is preferred
over one on the angle \(\theta\) itself, since its efficiency
depends less strongly on the boost and the mass of the
candidate~\cite{thesisVeerle}. The second criterion is that the
kinematic separation of the jets should satisfy
\(\Delta R = \sqrt{(\Delta\eta)^2+(\Delta\phi)^2} < 2.2\), where
\(\Delta\eta\) and \(\Delta\phi\) are the pseudorapidity and azimuthal
angle differences between the two jets, respectively. 
This reduces the tail in the dijet invariant mass
  distribution by suppressing the remaining back-to-back dijet
  background.

The overall efficiency to reconstruct and select displaced \vpion{}
decays in the simulated samples is summarized in
\cref{tab:efftable20112012} for the 2011 and 2012 data taking
conditions. A large part of the inefficiency is due to the detector
acceptance, which is about 13\% (8\%) and 6.5\% (5.5\%) for \vpion{}
particles with a lifetime of \SI{10}{\pico\second}
(\SI{100}{\pico\second}) and masses of \SI{25}{\GeVcc} and
\SI{50}{\GeVcc}, respectively.
Other important contributions are
  due to the selection on the displacement from the beamline,
  requirements on the minimum number of tracks forming the vertex, the
  material interaction veto, the reduction in \velo{} tracking
  efficiency at large displacements, and the jet selection~\cite{thesisPieter}.
The efficiency for long-lived particles decaying to \squark{}- and
\cquark{}-quark pairs is higher than for decays to \bquark{}-quark pairs
due to the larger number of tracks
originating directly from the \vpion{} decay vertex.

\ctable[ caption={Number of selected candidates per generated \Htopivpiv{} event (efficiency)
 in percent for different
  \decay{\vpion}{\quark{}\quarkbar{}}, \(\quark=\bquark, \cquark, \squark\)
 models for 2011 and 2012 data taking conditions, as derived from simulation.  The relative statistical
 uncertainty on the efficiency due to the limited size of the
 simulated sample is less than a few percent.
},label=tab:efftable20112012 ]
{
 l
 S[table-format=2.3(0)]
 S[table-format=2.3(0)]
 S[table-format=2.3(0)]
 S[table-format=2.3(0)]
 S[table-format=2.3(0)]
}
{}
{
\FL
&  & \multicolumn{2}{c}{2011} & \multicolumn{2}{c}{2012}
\ML
& \vpion{} mass & \SI{10}{\pico\second} & \SI{100}{\pico\second} & \SI{10}{\pico\second} & \SI{100}{\pico\second}
\ML
\decay{\vpion}{\bbbar{}} & {\SI{25}{\GeVcc}} & 0.45 & 0.097 & 0.46 & 0.111 \NN
\decay{\vpion}{\bbbar{}} & {\SI{35}{\GeVcc}} & 0.80 & 0.176 & 0.83 & 0.224 \NN
\decay{\vpion}{\bbbar{}} & {\SI{43}{\GeVcc}} & 0.73 & 0.190 & 0.77 & 0.222 \NN
\decay{\vpion}{\bbbar{}} & {\SI{50}{\GeVcc}} & 0.49 & 0.141 & 0.54 & 0.171 \NN
\decay{\vpion}{\ccbar{}} & {\SI{35}{\GeVcc}} & 1.35 &       & 1.35 & \NN
\decay{\vpion}{\ssbar{}} & {\SI{35}{\GeVcc}} & 1.30 &       & 1.19 & \LL
}

%% file: systematics.tex
\section{Systematic uncertainties}
\label{sec:systematics}

Systematic uncertainties on the efficiency
are obtained from studies of data-simulation differences in
control samples.
They are reported in
\cref{tab:combinedsyst2011,tab:combinedsyst2012}, for the 2011 and 2012
conditions, respectively, and discussed in more detail below. 
Uncertainties on the signal efficiency due to parton-density
distributions, the simulation of fragmentation and hadronization, and
the Higgs boson production cross-section and kinematics are not taken into account.

\ctable[
  label=tab:combinedsyst2011,
  caption={Overview of the contributions to the relative systematic uncertainty on the signal efficiency and luminosity (in percent) for different signal samples in 2011 conditions. The uncertainty on the total efficiency is obtained by summing the individual contributions in quadrature.}
  ]{lrrrrrrrrrr}
  {}
  {\FL
  \vpion{} mass (\si{\GeVcc})           &\multicolumn{2}{c}{25} &\multicolumn{2}{c}{35} &\multicolumn{2}{c}{43} &\multicolumn{2}{c}{50} &{35, \ccbar{}} &{35, \ssbar{}} \NN
  \vpion{} lifetime (\si{\pico\second}) &  \num{10} & \num{100} &  \num{10} & \num{100} &  \num{10} & \num{100} &  \num{10} & \num{100} &  \num{10}     &  \num{10}     \ML
  Tracking efficiency                   & \num{4.2} & \num{4.1} & \num{3.3} & \num{3.2} & \num{3.0} & \num{2.8} & \num{3.0} & \num{2.7} & \num{1.8}     & \num{1.7}     \NN
  Vertex finding                        & \num{3.8} & \num{4.2} & \num{3.3} & \num{3.9} & \num{2.8} & \num{3.7} & \num{3.7} & \num{2.6} & \num{2.9}     & \num{2.8}     \NN
  Jet reconstruction                    & \num{3.1} & \num{3.1} & \num{1.6} & \num{1.6} & \num{0.7} & \num{0.7} & \num{0.5} & \num{0.5} & \num{0.9}     & \num{1.0}     \NN
  Jet identification                    & \num{3.0} & \num{3.0} & \num{3.0} & \num{3.0} & \num{3.0} & \num{3.0} & \num{3.0} & \num{3.0} & \num{3.0}     & \num{3.0}     \NN
  Jet direction                         & \num{7.0} & \num{7.0} & \num{6.0} & \num{6.0} & \num{7.4} & \num{7.4} & \num{8.5} & \num{8.5} & \num{5.9}     & \num{5.7}     \NN
  \lone{}                               & \num{4.0} & \num{4.0} & \num{3.0} & \num{3.0} & \num{3.0} & \num{3.0} & \num{2.0} & \num{2.0} & \num{1.8}     & \num{2.1}     \NN
  \(N_{\text{\spd}}\)                   & \num{1.7} & \num{1.7} & \num{2.0} & \num{2.0} & \num{1.6} & \num{1.6} & \num{2.3} & \num{2.3} & \num{1.7}     & \num{1.6}     \NN
  \hltone{}                             & \num{2.0} & \num{2.0} & \num{2.0} & \num{2.0} & \num{2.0} & \num{2.0} & \num{2.0} & \num{2.0} & \num{2.0}     & \num{2.0}     \NN
  \hlttwo{}                             & \num{3.0} & \num{3.0} & \num{3.0} & \num{3.0} & \num{3.0} & \num{3.0} & \num{3.0} & \num{3.0} & \num{3.0}     & \num{3.0}     \ML
  Total efficiency                      &\num{11.5} &\num{11.6} & \num{9.8} &\num{10.0} &\num{10.3} &\num{10.5} &\num{11.2} &\num{10.9} & \num{8.7}     & \num{8.6}     \ML
  Luminosity                            & \num{1.7} & \num{1.7} & \num{1.7} & \num{1.7} & \num{1.7} & \num{1.7} & \num{1.7} & \num{1.7} & \num{1.7}     & \num{1.7}     \LL
  }

\ctable[
  label=tab:combinedsyst2012,
  caption={Overview of the contributions to the relative systematic uncertainty
    on the signal efficiency and luminosity (in percent) for different
    signal samples in 2012 conditions.  The uncertainty on the total efficiency is obtained by summing the individual contributions in quadrature.}
  ]{lrrrrrrrrrr}
  {}
  {\FL
  \vpion{} mass (\si{\GeVcc})           &\multicolumn{2}{c}{25} &\multicolumn{2}{c}{35} &\multicolumn{2}{c}{43} &\multicolumn{2}{c}{50} &{35, \ccbar{}} &{35, \ssbar{}} \NN
  \vpion{} lifetime (\si{\pico\second}) &  \num{10} & \num{100} &  \num{10} & \num{100} &  \num{10} & \num{100} &  \num{10} & \num{100} &  \num{10}     &  \num{10}     \ML
  Tracking efficiency                   & \num{3.1} & \num{2.8} & \num{2.4} & \num{2.4} & \num{2.2} & \num{2.1} & \num{2.0} & \num{1.7} & \num{1.2}     & \num{1.1}     \NN 
  Vertex finding                        & \num{4.2} & \num{4.5} & \num{3.8} & \num{4.4} & \num{3.4} & \num{4.1} & \num{3.1} & \num{3.9} & \num{3.4}     & \num{3.5}     \NN
  Jet reconstruction                    & \num{2.7} & \num{2.7} & \num{1.1} & \num{1.1} & \num{0.7} & \num{0.7} & \num{0.3} & \num{0.3} & \num{0.9}     & \num{1.0}     \NN
  Jet identification                    & \num{3.0} & \num{3.0} & \num{3.0} & \num{3.0} & \num{3.0} & \num{3.0} & \num{3.0} & \num{3.0} & \num{3.0}     & \num{3.0}     \NN
  Jet direction                         & \num{5.8} & \num{5.8} & \num{5.3} & \num{5.3} & \num{6.1} & \num{6.1} & \num{7.9} & \num{7.9} & \num{5.3}     & \num{5.8}     \NN
  \lone{}                               & \num{4.0} & \num{4.0} & \num{2.5} & \num{2.5} & \num{2.0} & \num{2.0} & \num{2.0} & \num{2.0} & \num{2.0}     & \num{2.0}     \NN
  \(N_{\text{\spd}}\)                   & \num{2.2} & \num{2.2} & \num{2.5} & \num{2.5} & \num{2.5} & \num{2.5} & \num{2.5} & \num{2.5} & \num{2.4}     & \num{2.1}     \NN
  \hltone{}                             & \num{2.0} & \num{2.0} & \num{2.0} & \num{2.0} & \num{2.0} & \num{2.0} & \num{2.0} & \num{2.0} & \num{2.0}     & \num{2.0}     \NN
  \hlttwo{}                             & \num{3.0} & \num{3.0} & \num{3.0} & \num{3.0} & \num{3.0} & \num{3.0} & \num{3.0} & \num{3.0} & \num{3.0}     & \num{3.0}     \ML
  Total efficiency                      &\num{10.5} &\num{10.6} & \num{9.2} & \num{9.4} & \num{9.1} & \num{9.5} &\num{10.4} &\num{10.6} & \num{8.6}     & \num{8.9}     \ML
  Luminosity                            & \num{1.2} & \num{1.2} & \num{1.2} & \num{1.2} & \num{1.2} & \num{1.2} & \num{1.2} & \num{1.2} & \num{1.2}     & \num{1.2}     \LL
}

The vertex reconstruction efficiency can be split into two parts,
namely the track reconstruction efficiency and the vertex finding
efficiency.  The track reconstruction efficiency
is described by the simulation to within a few percent, including for
highly displaced and low-momentum
tracks~\cite{LHCb-DP-2013-002,LHCb-PAPER-2010-001,LHCb-PAPER-2011-005}.
The effect of a systematic change in this efficiency is studied by
randomly removing 2\% of the signal tracks and
reapplying all selection criteria.

The vertex finding algorithm is not fully efficient even if all tracks
are reconstructed. In particular, the efficiency to find a low-multiplicity secondary
vertex is reduced in the proximity of a high-multiplicity PV.
The effect is studied in data and simulation using exclusively
reconstructed \BdToJPsiKst{} decays, which can be selected with high
purity without tight requirements on the vertex.
The efficiency for the displaced vertex reconstruction algorithm to
find the \Bz{} candidate is measured as a function of the displacement
\Rxy{} in data and simulation~\cite{thesisPieter}.
The difference, weighted by the \Rxy{} distribution of the signal candidates, is used to derive a systematic uncertainty.

Systematic uncertainties related to the jet reconstruction can be
introduced in two ways: through differences between data and
simulation in the jet reconstruction efficiency and through
differences between data and simulation in the resolution on the jet
energy and direction, which enter the dijet candidate kinematic and
\(m/m_\text{corr}\) selection and the dijet invariant mass shape.  The
jet reconstruction efficiency has been studied previously in
measurements of the \(\Z+\text{jet}\) and \(\Z+\text{\bquark{}-jet}\) cross-sections and was found to be
consistent between data and simulation~\cite{LHCb-PAPER-2013-058,LHCb-PAPER-2014-055}.
The \(\decay{\Z}{\mumu}+\text{jet}\) sample is used to study
  jet-related systematic effects for this analysis as well. To mimic
  the selection of the particle-flow inputs, the PV associated to the
  $\Z$ is used as a proxy for the displaced vertex.

The difference between data and simulation with the largest impact on
the jet reconstruction efficiency is the energy response to low-\pt{} jets,
close to the threshold of \SI{5}{\GeVc}. The sensitivity to a
different energy response in data and simulation is evaluated by
increasing the minimum jet \pt{} for candidates passing the full
offline selection by 10\%, which is the uncertainty on
the jet energy scale.  The change in the overall selection efficiency
is assigned as a systematic uncertainty.
By replacing the jet identification criteria with a requirement on the
\pt{} balance between the leading jet and the \Z{} boson, the
\Ztomumu{} sample can also be used to study the difference in jet
identification efficiency between data and simulation. 
No difference larger than 3\% relative is seen, which is assigned as a
systematic uncertainty.

To validate the simulation of the jet-direction
  resolution the jet-direction is estimated separately with the
  charged and neutral components of the jet in \(\Z{}+\text{jet}\)
  events.  The distribution of the charged-neutral difference in the
  estimated direction is found to be consistent between data and
  simulation for both the \(\eta\) and the \(\phi\) projection, and
  across the full range of \pt{}. To quantify the effect on the
  \vpion{} signal efficiency, an additional smearing to the
  jet-direction is applied to jets of selected candidates in the
  simulation. The jet angles with respect to the beam direction are
  smeared independently in the horizontal and vertical planes by about
  one third of the resolution, which is the largest value
  compatible with the comparison of data and simulation in  \(\Z{}+\text{jet}\)
  events.

The systematic uncertainty related to the \lone{} trigger selection
consists of two parts, due to differences in the \lone{} calorimeter
trigger response between data and simulation, and due to the difference between data and simulation in
the distribution of the \spd{} hit multiplicity \(N_\text{\spd}\). The first is evaluated
by studying the \lone{} calorimeter trigger response on jets reconstructed in \(\Z{}+\text{jet}\)
events, where the trigger decision is made based on the \Ztomumu{} decay products, and is independent of the jet.
The observed data-simulation differences are propagated to the \vpion{} reconstruction efficiency and correspond to systematic uncertainties of 2--4\%, depending on the \vpion{} mass.
Jets in \(\Z{}+\text{jet}\) events are mostly
  light-quark jets, while our benchmark signal decays to $b$ quarks. 
  It is found in simulated events that the efficiency of the
  \lone{} calorimeter trigger is practically independent of jet
  flavour. A small fraction of $b$-quark jets is triggered
  exclusively by the \lone{} muon trigger, which is well modelled in
  the simulation.

The second part of the \lone{} systematic uncertainty arises because the \spd{} multiplicity is not well described in the simulation.
This effect is studied with a \Ztomumu{} sample triggered by the
dimuon \lone{} selection, which applies only a loose selection on this
quantity. An efficiency
correction is derived, which is about 90\% for 2011 data, and about 85\% for 2012 data,
with an uncertainty of 2--3\%.
The difference in the correction between the different \vpion{} models is smaller than the systematic variation.
This correction is applied to the overall detection efficiency derived from the simulation and the uncertainty is taken as a systematic uncertainty.

The differences between data and simulation in the \hltone{} selection are dominated by the track
reconstruction efficiency, which was discussed above, and additional
track quality criteria. One such difference is due to a requirement
on the number of \velo{} hits for displaced tracks. It is
characterized using \BdToJPsiKst{} decays selected with triggers that do not apply such a requirement. For this sample
the selection efficiency was found to be 2\% higher in
data than in simulated events, which is assigned as a systematic uncertainty. For \vpion{} decays the final-state
track multiplicity is larger, which dilutes effects due to a
mismodelling of the single-track efficiency.

The main source of systematic uncertainty in the \hlttwo{} selection is the vertex
reconstruction efficiency, which was discussed above. The
efficiency of the topological \B{} trigger, which is relevant for a subset of
the candidates, is accurately described in
simulation. It is measured as a function of \Rxy{} in
data and simulation using \BdToJPsiKst{} candidates that are selected by a different, dimuon-based, trigger criterion. A
maximum difference of 2--3\% is observed, which is
assigned as a systematic uncertainty.

%% file: results.tex
\section{Results}
\label{sec:results}

\begin{figure}[tbp]
  \begin{subfigure}{.5\textwidth}%
    \centering%
    \includegraphics[width=\linewidth]{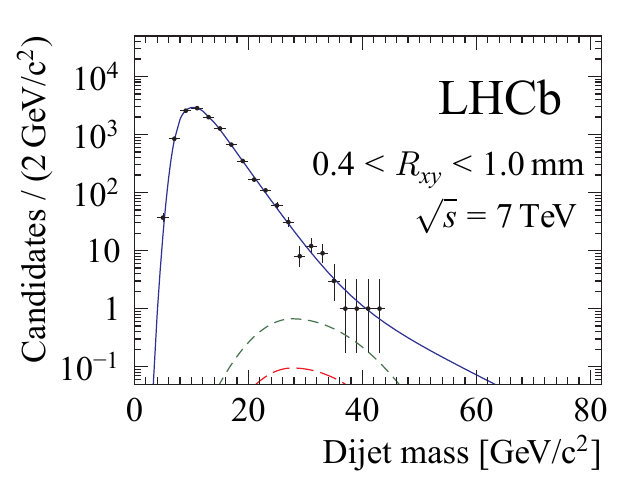}%
    \phantomcaption\label{fig:allInvMassFIts_M35_2011_bin1}%
  \end{subfigure}%
  \hfill%
  \begin{subfigure}{.5\textwidth}%
    \centering%
    \includegraphics[width=\linewidth]{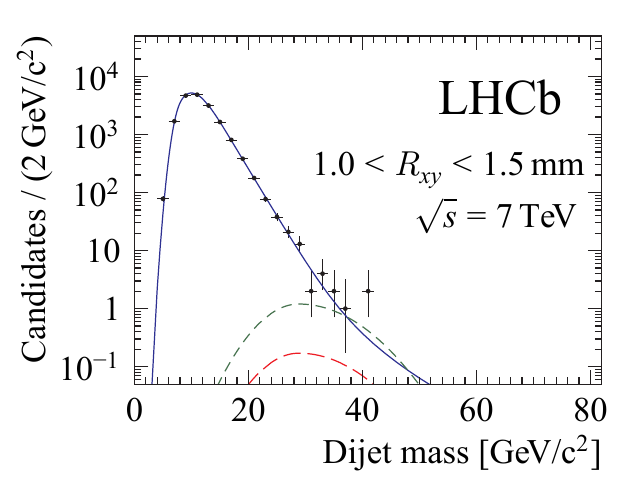}%
    \phantomcaption\label{fig:allInvMassFIts_M35_2011_bin2}%
  \end{subfigure}%

  \begin{subfigure}{.5\textwidth}%
    \centering%
    \includegraphics[width=\linewidth]{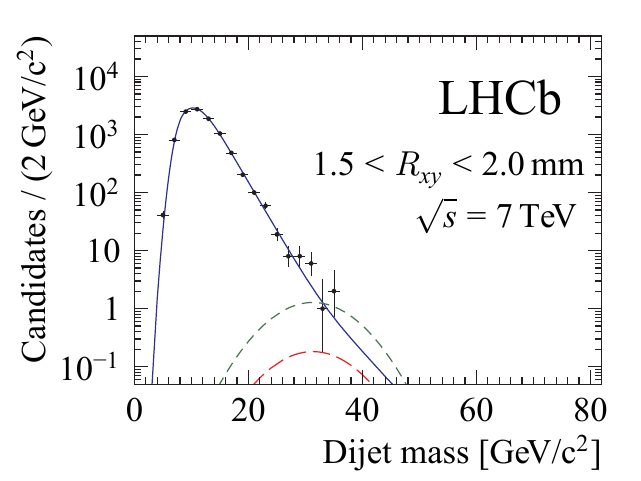}%
    \phantomcaption\label{fig:allInvMassFIts_M35_2011_bin3}%
  \end{subfigure}%
  \hfill%
  \begin{subfigure}{.5\textwidth}%
    \centering%
    \includegraphics[width=\linewidth]{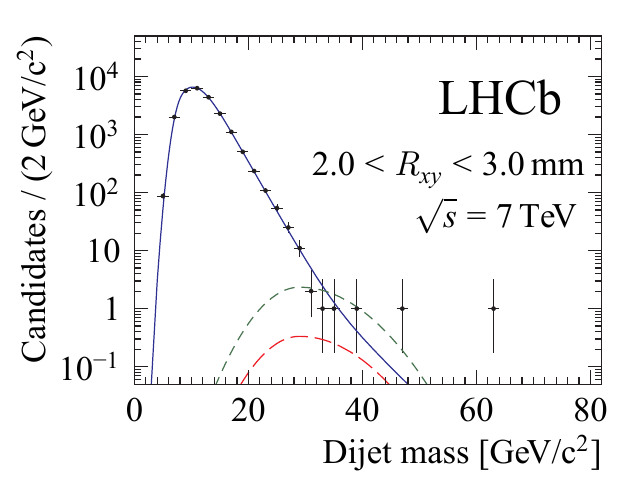}%
    \phantomcaption\label{fig:allInvMassFIts_M35_2011_bin4}%
  \end{subfigure}%

  \begin{subfigure}{.5\textwidth}%
    \centering%
    \includegraphics[width=\linewidth]{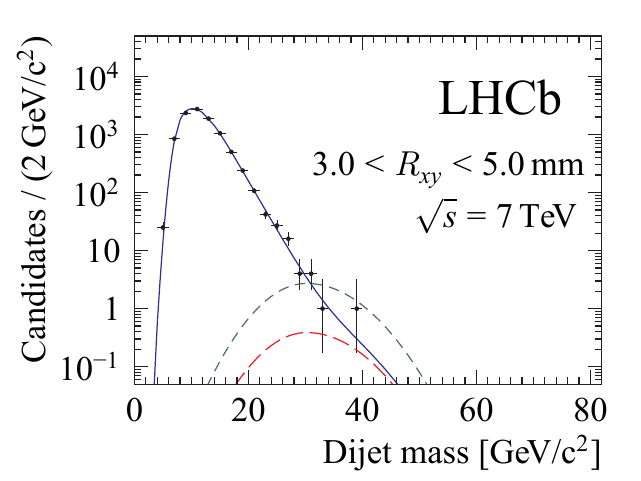}%
    \phantomcaption\label{fig:allInvMassFIts_M35_2011_bin5}%
  \end{subfigure}%
  \hfill%
  \begin{subfigure}{.5\textwidth}%
    \centering%
    \includegraphics[width=\linewidth]{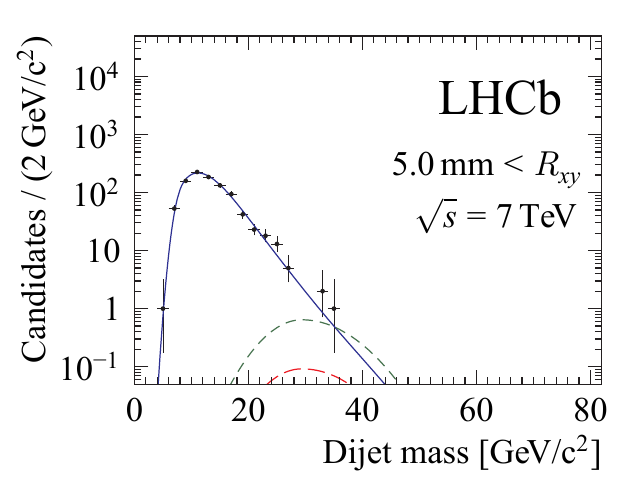}%
    \phantomcaption\label{fig:allInvMassFIts_M35_2011_bin6}%
  \end{subfigure}%
  \caption{Dijet invariant mass distribution in the different \Rxy{} bins, for the 2011 data sample.
  For illustration, the best fit with a signal \vpion{} model with mass \SI{35}{\GeVcc} and lifetime \SI{10}{\pico\second} is overlaid.
  The solid blue line indicates the total background model,
  the short-dashed green line indicates the signal model for signal strength $\mu=1$, and the long-dashed red line
  indicates the best-fit signal strength.}
  \label{fig:allInvMassFIts_M35_2011}
\end{figure}

\begin{figure}[htbp]
  \begin{subfigure}{.5\textwidth}%
    \centering%
    \includegraphics[width=\linewidth]{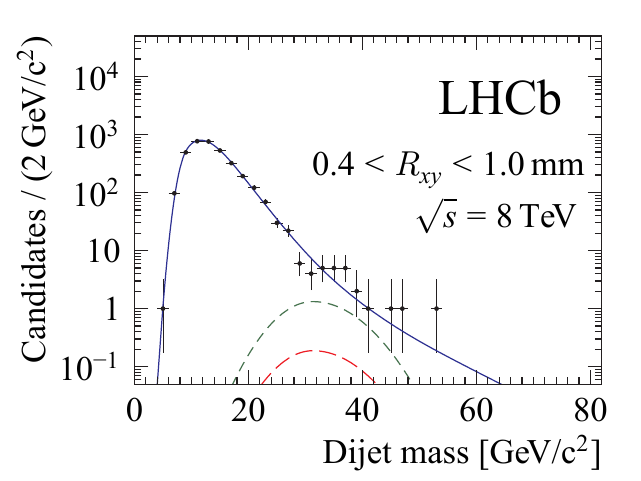}%
    \phantomcaption\label{fig:allInvMassFIts_M35_2012_bin1}%
  \end{subfigure}%
  \hfill%
  \begin{subfigure}{.5\textwidth}%
    \centering%
    \includegraphics[width=\linewidth]{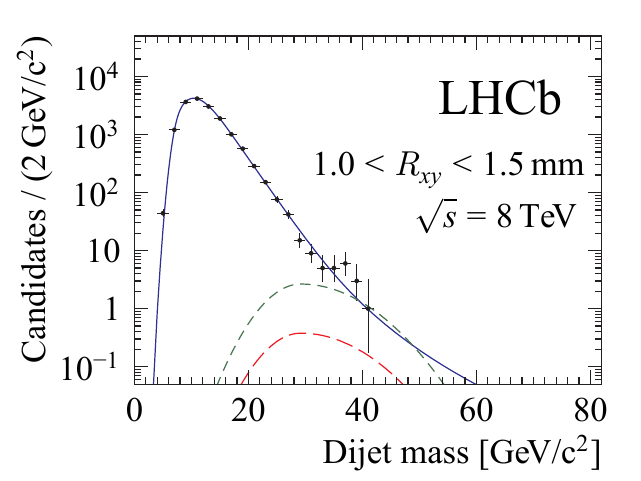}%
    \phantomcaption\label{fig:allInvMassFIts_M35_2012_bin2}%
  \end{subfigure}%

  \begin{subfigure}{.5\textwidth}%
    \centering%
    \includegraphics[width=\linewidth]{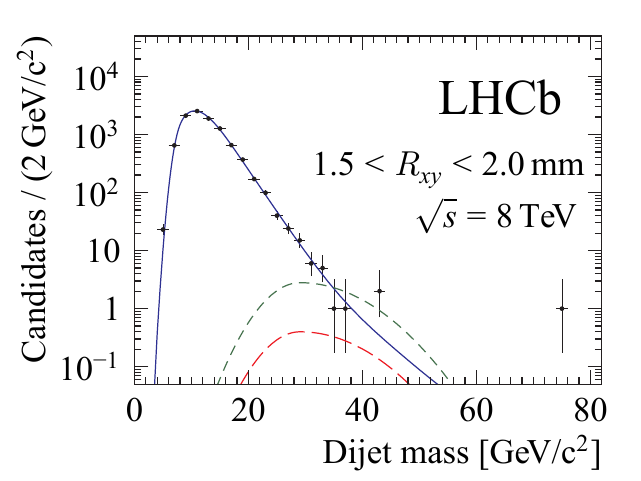}%
    \phantomcaption\label{fig:allInvMassFIts_M35_2012_bin3}%
  \end{subfigure}%
  \hfill%
  \begin{subfigure}{.5\textwidth}%
    \centering%
    \includegraphics[width=\linewidth]{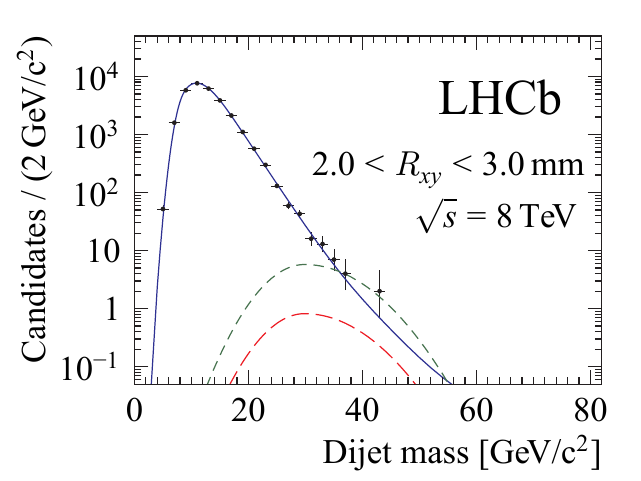}%
    \phantomcaption\label{fig:allInvMassFIts_M35_2012_bin4}%
  \end{subfigure}%

  \begin{subfigure}{.5\textwidth}%
    \centering%
    \includegraphics[width=\linewidth]{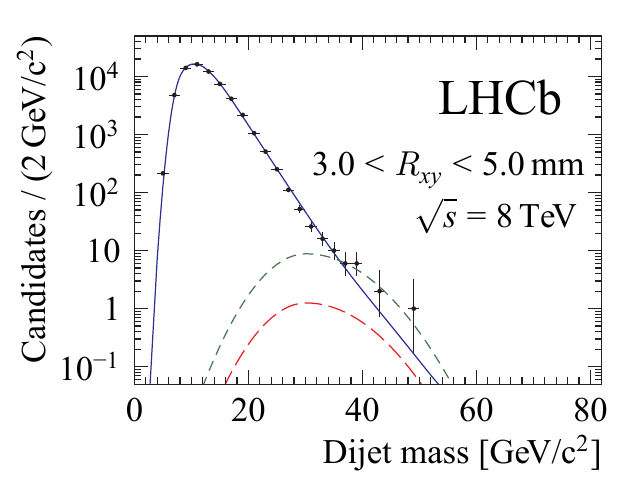}%
    \phantomcaption\label{fig:allInvMassFIts_M35_2012_bin5}%
  \end{subfigure}%
  \hfill%
  \begin{subfigure}{.5\textwidth}%
    \centering%
    \includegraphics[width=\linewidth]{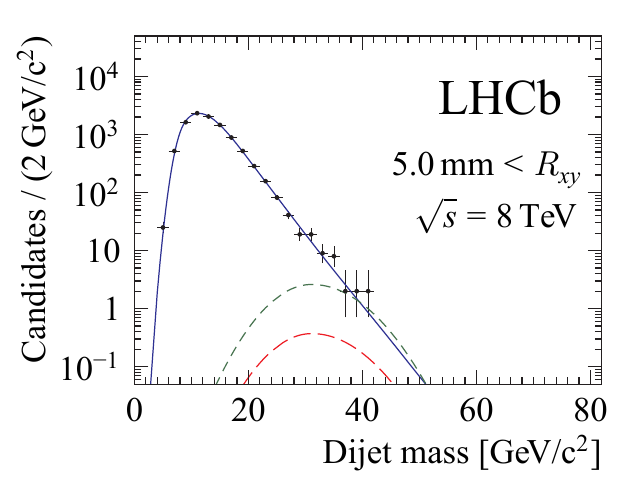}%
    \phantomcaption\label{fig:allInvMassFIts_M35_2012_bin6}%
  \end{subfigure}%
  \caption{Dijet invariant mass distribution in the different \Rxy{} bins, for the 2012 data sample.
  For illustration, the best fit with a signal \vpion{} model with mass \SI{35}{\GeVcc} and lifetime \SI{10}{\pico\second} is overlaid.
  The solid blue line indicates the total background model,
  the short-dashed green line indicates the signal model for signal strength $\mu=1$, and the long-dashed red line
  indicates the best-fit signal strength.}
  \label{fig:allInvMassFIts_M35_2012}
\end{figure}

Constraints on the presence of a signal are derived from a fit to the
dijet invariant mass distributions, shown in
\cref{fig:allInvMassFIts_M35_2011,fig:allInvMassFIts_M35_2012}. To
take advantage of the difference in the \Rxy{} distribution for
background and signal, the data are divided into six $\Rxy$
bins. The data are further split according to data taking year to
account for differences in running conditions and
Higgs boson production cross-section. The signal efficiency for each \Rxy{} bin is
obtained from the simulated samples with \vpion{} lifetimes of
\SI{10}{\pico\second} and \SI{100}{\pico\second}, with the decay time
distributions reweighted to mimic other lifetime hypotheses as
needed.

Results are presented as upper limits on the signal strength
\(\mu \equiv (\sigma/\sigma_{gg\rightarrow\Higgs{}}^{SM})\cdot
\mathcal{B}(\Htopivpiv{})\), where $\sigma$ is the excluded signal
cross-section, $\sigma_{gg\rightarrow\Higgs{}}^{SM}$ is the SM Higgs
boson production cross-section via the gluon fusion process and
$\mathcal{B}(\Htopivpiv{})$ is the branching fraction of the Higgs
boson decay to $\vpion{}$ particles. The branching fraction \(\BR_{\quark\quarkbar}\) of the
\vpion{} particle to the \qqbar{} final state (with $\qqbar=\bbbar$, $\ccbar$
or $\ssbar$ depending on the final state under study) is assumed to be
100\%. If the decay width of the \vpion{} particle is dominated by other decays
than that under study, the limits scale as \(1/(\BR_{\quark{}\quarkbar{}}(2-\BR_{\quark{}\quarkbar{}}))\).
The Higgs boson production cross-section is assumed to be
\SI{15.11}{\pico\barn} at \SI{7}{\tera\electronvolt} and
\SI{19.24}{\pico\barn} at
\SI{8}{\tera\electronvolt}~\autocite{Heinemeyer:2013tqa}.

The \(\text{CL}_s\) method~\cite{Read:CLs2002} is used to determine
upper limits. The profile likelihood ratio
\(
  q_\text{PLL}^\mu = {L(\mu,\hat{\theta}(\mu))}/{L(\hat{\mu},\hat{\theta})}
\)
is chosen as a test statistic, where \(L(\mu,\theta)\) denotes the
likelihood as a function of $\mu$ and a set of nuisance parameters
$\theta$, which are also extracted from the data;
\(L(\mu,\hat{\theta}(\mu))\) is the maximum likelihood for a
hypothesized value of \(\mu\) and \(L(\hat{\mu},\hat{\theta})\) is the
global maximum likelihood.  To estimate the sensitivity of the
analysis and the significance of a potential signal, the expected
upper limit quantiles in the case of zero signal are also evaluated. 

For each value of $\mu$ and $\theta$ the likelihood is evaluated as
\(L(\mu,\theta) = \prod_{i} P(x_i;\mu,\theta)\),
where $P$ is the probability density for event $i$ and the product
runs over all selected events. The observables $x_i$ for each candidate include
the dijet mass, \Rxy{} bin and data taking year. For
each \Rxy{} bin and data taking year, the invariant mass distribution is modelled by
the sum of background and signal components. The distribution for the
signal is modelled as a Gaussian distribution whose parameters are
obtained from fully simulated signal events.
For the background distribution an empirical model, outlined below, is
adopted.

Background candidates can be categorized into two
contributions.  The first category is mostly due to the combination of a
heavy-flavour decay vertex or an interaction with detector material
with particles from a primary interaction. This contribution has a
steeply decreasing invariant mass spectrum. Following the approach
in Ref.~\cite{LHCb-PAPER-2014-062}, the distribution is modelled by the
convolution of a falling exponential distribution with a bifurcated
Gaussian. All parameters of this background model are free to vary in
the fit.

The second category is due to Standard Model dijet events. These
events have candidates with jets that are approximately back-to-back
in the transverse plane. It is suppressed by the selection on the
dijet opening angle $\Delta R$.  Its remaining contribution has a less
steeply falling mass spectrum.  It is described in the fit with a
similar functional shape as for the first category, but with the
parameters and the relative yields in the different bins fixed from a
fit to the invariant mass distribution of candidates that fail the
$\Delta R$ requirement. In the final fit only the total normalization
of this component is varied. The second component is new compared to
the model used for the previous
analysis~\cite{LHCb-PAPER-2014-062}. It leads to a better description
of the high-mass tail, at the expense of one extra fit parameter for each
data taking year.
It was found that the result of the fit is not sensitive
to the exact $\Delta R$ requirement used to select the events for this component.

All parameters of the fit to the invariant mass distribution are
allowed to float independently in each bin, except for the following
nuisance parameters: the dijet invariant mass scale, the overall signal efficiency,
and the normalization for the second background contribution.  All
relevant systematic uncertainties are incorporated in the fit model:
the overall uncertainty on the efficiency, as described in
\cref{sec:systematics}, the uncertainty on the dijet invariant mass
scale, and the uncertainties on the shape parameters and relative
normalisation arising from the finite size of the simulated signal
samples. Gaussian constraints on
these parameters are added to the likelihood.

Alternatives have been considered for the background
  mass model, in particular with an additional less steeply falling exponential to
  describe the tail. With these models the estimated background yield
  at higher mass is similar or larger than with the nominal background
  model, leading to tighter limits on the signal. As the nominal model
  gives the most conservative limit, no additional systematic
  uncertainty is assigned for background modeling.

\begin{figure}[htbp]
  \begin{subfigure}{.5\textwidth}%
    \centering%
    \includegraphics[width=\linewidth]{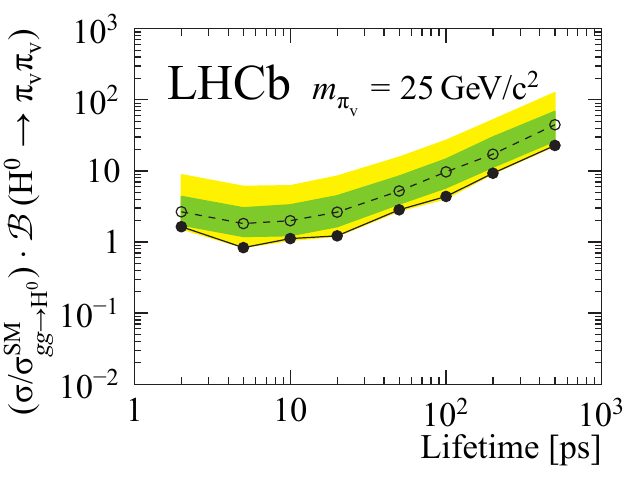}%
    \phantomcaption\label{fig:exclusion_M25_ctau}%
  \end{subfigure}%
  \hfill%
  \begin{subfigure}{.5\textwidth}%
    \centering%
    \includegraphics[width=\linewidth]{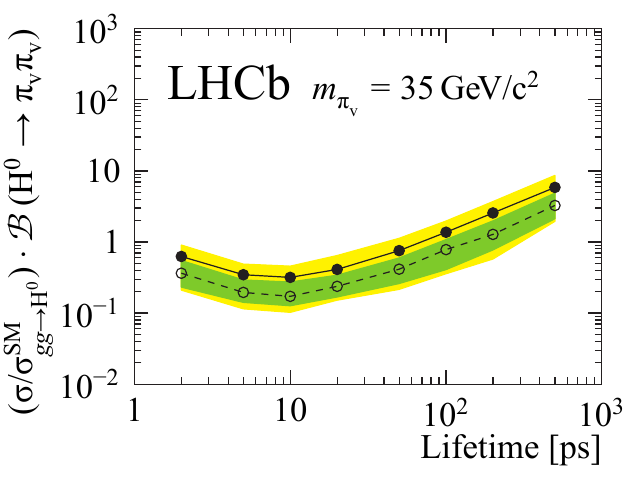}%
    \phantomcaption\label{fig:exclusion_M35_ctau}%
  \end{subfigure}%

  \begin{subfigure}{.5\textwidth}%
    \centering%
    \includegraphics[width=\linewidth]{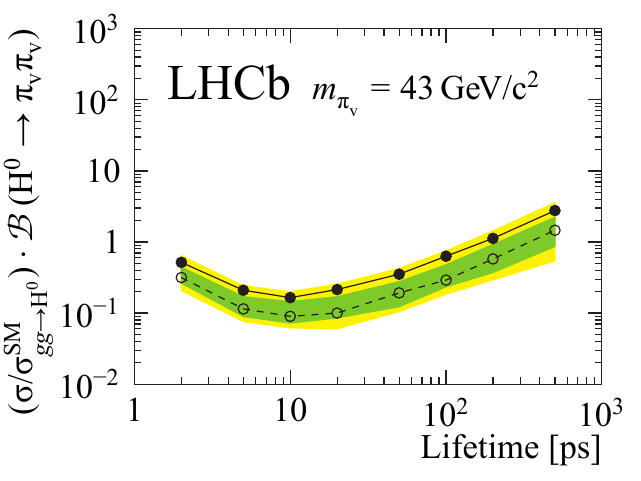}%
    \phantomcaption\label{fig:exclusion_M43_ctau}%
  \end{subfigure}%
  \hfill%
  \begin{subfigure}{.5\textwidth}%
    \centering%
    \includegraphics[width=\linewidth]{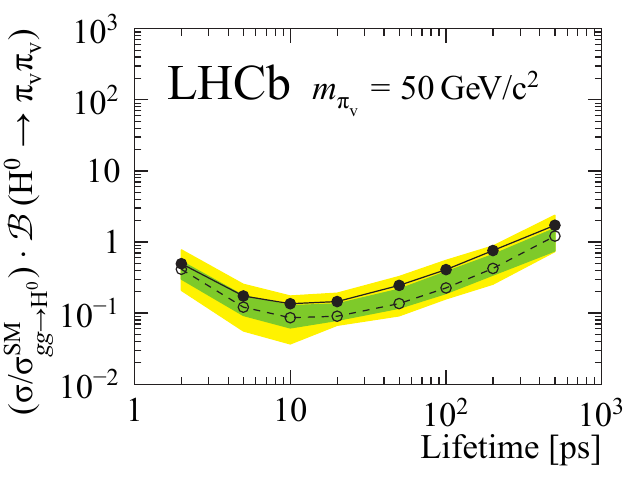}%
    \phantomcaption\label{fig:exclusion_M50_ctau}%
  \end{subfigure}%

  \begin{subfigure}{.5\textwidth}%
    \centering%
    \includegraphics[width=\linewidth]{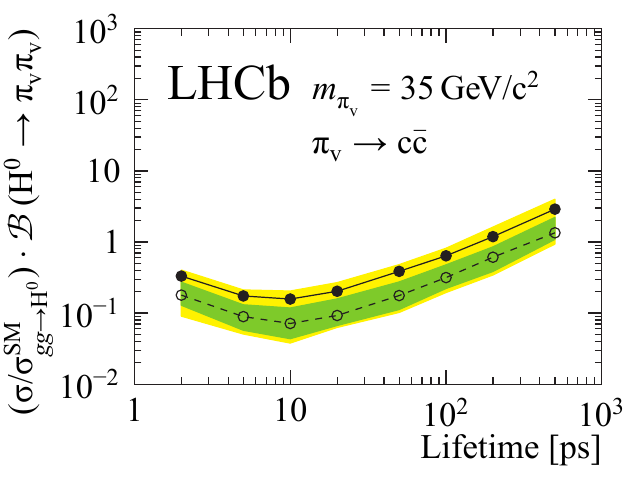}%
    \phantomcaption\label{fig:exclusion_CC_ctau}%
  \end{subfigure}%
  \hfill%
  \begin{subfigure}{.5\textwidth}%
    \centering%
    \includegraphics[width=\linewidth]{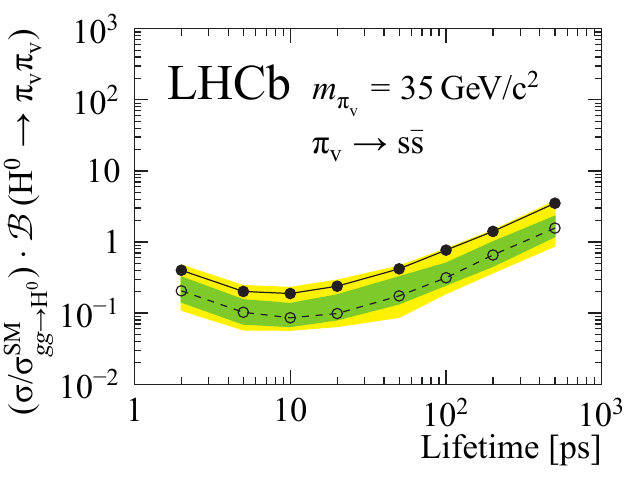}%
    \phantomcaption\label{fig:exclusion_SS_ctau}%
  \end{subfigure}%
  \caption{Expected (open circles and dotted line) and observed
    (filled circles and solid line) upper limit versus lifetime for
    different \vpion{} masses and decay modes. The green (dark) and
    yellow (light) bands indicate the quantiles of the expected upper limit corresponding to \(\pm 1\sigma\)
    and \(\pm 2\sigma\) for a Gaussian distribution.
    The decay \decay{\vpion{}}{\bbbar{}} is assumed, unless specified otherwise.}
  \label{fig:exclusionbrazilvstau_2ndexp}
\end{figure}

\begin{figure}
  \includegraphics[width=\fullfigwidth]{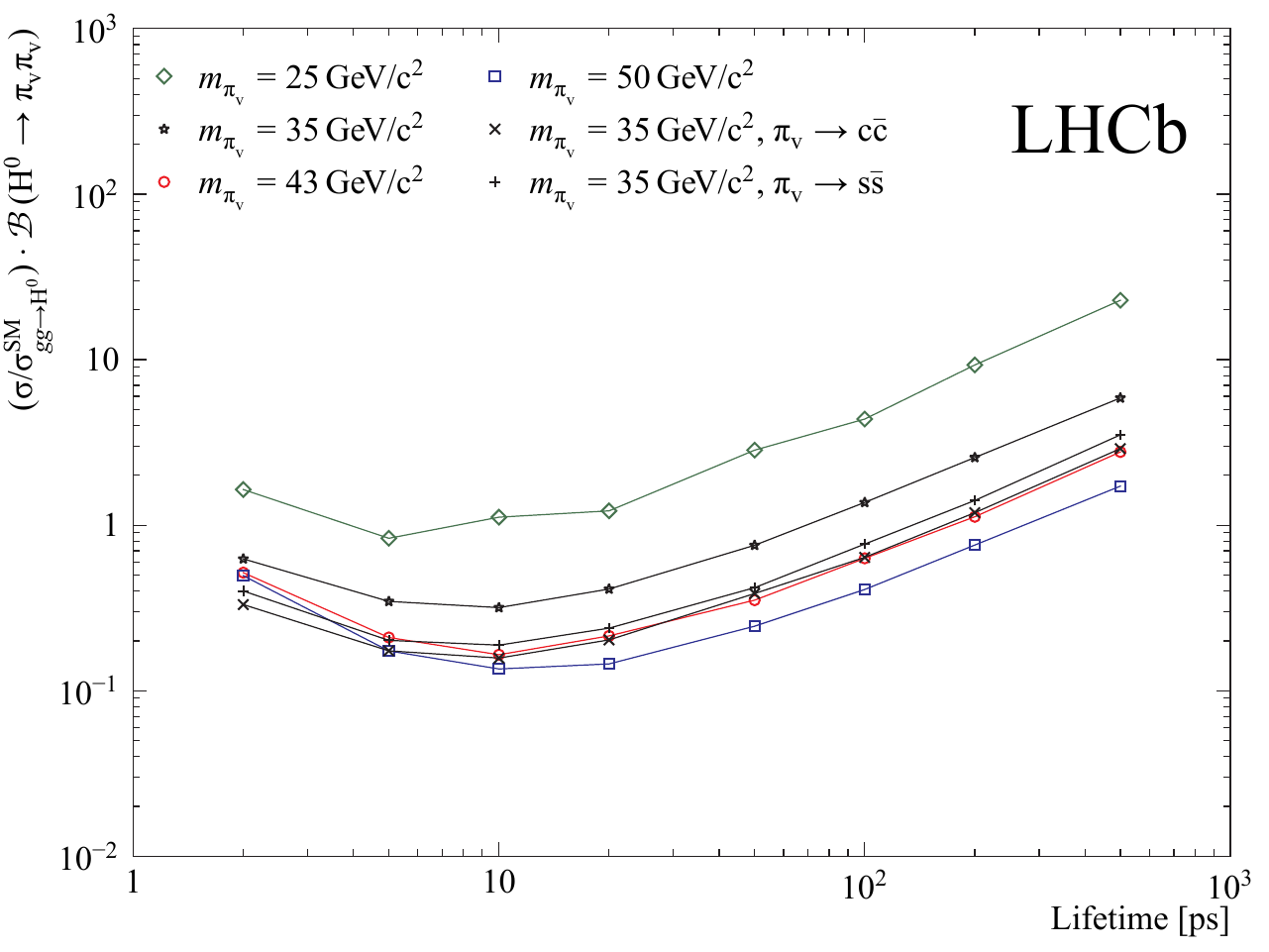}%
  \caption{Observed upper limit versus lifetime for different \vpion{} masses and decay modes.
    The decay \decay{\vpion{}}{\bbbar{}} is assumed, unless specified otherwise.}
  \label{fig:exclusionvstau_all}
\end{figure}

There is no significant excess of signal  in the
data. Upper limits at 95\% confidence level (CL) as a function of lifetime for
hidden-valley models with different \vpion{} mass and decay mode are shown in
\cref{fig:exclusionbrazilvstau_2ndexp} and summarized in
\cref{tab:observedlimits} and \cref{fig:exclusionvstau_all}.
The best sensitivity is obtained for a mass of about \SI{50}{\GeVcc}
and a lifetime of about \SI{10}{\pico\second}. The main
improvements with respect to the previous result~\cite{LHCb-PAPER-2014-062}
are due to the enlarged data sample, the improved trigger selections,
and the addition of the \Rxy{} bin above \SI{5}{\milli\meter}, which
contributes to the increased sensitivity at larger lifetimes.

\input{observedLimits.dat}

\clearpage
\section{Conclusion}
\label{sec:discussionandoutlook}

Results have been presented from a search for long-lived particles with a
mass in the range \SIrange{25}{50}{\GeVcc} and a lifetime between $2$ and 
\SI{500}{\pico\second}. The particles are
assumed to be pair-produced in the decay of a \SI{125}{\GeVcc}
Standard-Model-like Higgs boson and to decay into two jets. Besides
decays to \bbbar{}, which are the best motivated in the context
of hidden-valley models~\cite{StrasslerZurekHV,StrasslerZurekHVHiggs},
also decays to \ccbar{} and \ssbar{} quark pairs are considered. No evidence for
so far unknown long-lived particles is observed and limits are set as a function of
mass and lifetime. 
These measurements complement other constraints on this production
model at the LHC~\cite{CMSDVDiJet2012,ATLASDVEXO2012} by placing
stronger constraints at small masses and lifetimes.

%% file: observedLimits.dat
\ctable[
   caption={Observed 95\% CL signal strength (\(\mu\)) upper limits for different \vpion{} models.}
  ,label={tab:observedlimits}
]
{
  l
  r
  r
  r
  r
  r
  r
  r
  r
}
{}
{
\toprule
 & \multicolumn{8}{c}{\vpion{} lifetime (\si{\pico\second})} \NN
\vpion{} mass                       & {\num{2}} & {\num{5}} & {\num{10}} & {\num{20}} & {\num{50}} & {\num{100}} & {\num{200}} & {\num{500}}
\ML
                                            \SI{25}{\GeVcc} &      1.64 &      0.83 &       1.12 &       1.22 &       2.84 &        4.37 &        9.28 &       22.82 \NN
                                            \SI{35}{\GeVcc} &      0.63 &      0.35 &       0.32 &       0.41 &       0.76 &        1.37 &        2.56 &        5.86 \NN
                                            \SI{43}{\GeVcc} &      0.52 &      0.21 &       0.16 &       0.21 &       0.35 &        0.63 &        1.12 &        2.77 \NN
                                            \SI{50}{\GeVcc} &      0.50 &      0.17 &       0.14 &       0.15 &       0.25 &        0.41 &        0.76 &        1.72 \NN
  \SI{35}{\GeVcc}, \(\vpion{}\to \cquark{}\bar{\cquark{}}\) &      0.33 &      0.17 &       0.16 &       0.20 &       0.39 &        0.64 &        1.19 &        2.90 \NN
  \SI{35}{\GeVcc}, \(\vpion{}\to \squark{}\bar{\squark{}}\) &      0.40 &      0.20 &       0.19 &       0.24 &       0.42 &        0.77 &        1.41 &        3.51
\LL
}

%% file: acknowledgements.tex
\clearpage
\section*{Acknowledgements}

\noindent We express our gratitude to our colleagues in the CERN
accelerator departments for the excellent performance of the LHC. We
thank the technical and administrative staff at the LHCb
institutes. We acknowledge support from CERN and from the national
agencies: CAPES, CNPq, FAPERJ and FINEP (Brazil); MOST and NSFC (China);
CNRS/IN2P3 (France); BMBF, DFG and MPG (Germany); INFN (Italy); 
NWO (The Netherlands); MNiSW and NCN (Poland); MEN/IFA (Romania); 
MinES and FASO (Russia); MinECo (Spain); SNSF and SER (Switzerland); 
NASU (Ukraine); STFC (United Kingdom); NSF (USA).
We acknowledge the computing resources that are provided by CERN, IN2P3 (France), KIT and DESY (Germany), INFN (Italy), SURF (The Netherlands), PIC (Spain), GridPP (United Kingdom), RRCKI and Yandex LLC (Russia), CSCS (Switzerland), IFIN-HH (Romania), CBPF (Brazil), PL-GRID (Poland) and OSC (USA). We are indebted to the communities behind the multiple open 
source software packages on which we depend.
Individual groups or members have received support from AvH Foundation (Germany),
EPLANET, Marie Sk\l{}odowska-Curie Actions and ERC (European Union), 
Conseil G\'{e}n\'{e}ral de Haute-Savoie, Labex ENIGMASS and OCEVU, 
R\'{e}gion Auvergne (France), RFBR and Yandex LLC (Russia), GVA, XuntaGal and GENCAT (Spain), Herchel Smith Fund, The Royal Society, Royal Commission for the Exhibition of 1851 and the Leverhulme Trust (United Kingdom).

%% file: LHCb_Authorship_flat_20-Dec-2016.tex
\centerline{\large\bf LHCb collaboration}
\begin{flushleft}
\small
R.~Aaij$^{40}$,
B.~Adeva$^{39}$,
M.~Adinolfi$^{48}$,
Z.~Ajaltouni$^{5}$,
S.~Akar$^{59}$,
J.~Albrecht$^{10}$,
F.~Alessio$^{40}$,
M.~Alexander$^{53}$,
S.~Ali$^{43}$,
G.~Alkhazov$^{31}$
% ...

{\footnotesize \it
% ...
$ ^{5}$Clermont Universit{\'e}, Universit{\'e} Blaise Pascal, CNRS/IN2P3, LPC, Clermont-Ferrand, France\\
% ...
$ ^{10}$Fakult{\"a}t Physik, Technische Universit{\"a}t Dortmund, Dortmund, Germany\\
% ...
$ ^{31}$Petersburg Nuclear Physics Institute (PNPI), Gatchina, Russia\\
% ...
$ ^{39}$Universidad de Santiago de Compostela, Santiago de Compostela, Spain\\
$ ^{40}$European Organization for Nuclear Research (CERN), Geneva, Switzerland\\
% ...
$ ^{43}$Nikhef National Institute for Subatomic Physics, Amsterdam, The Netherlands\\
% ...
$ ^{48}$H.H. Wills Physics Laboratory, University of Bristol, Bristol, United Kingdom\\
% ...
$ ^{53}$School of Physics and Astronomy, University of Glasgow, Glasgow, United Kingdom\\
% ...
$ ^{59}$University of Cincinnati, Cincinnati, OH, United States\\
% ...
}
\end{flushleft}